\documentclass[pdflatex,sn-mathphys-num]{sn-jnl}

\usepackage{graphicx}%
\usepackage{multirow}%
\usepackage{amsmath,amssymb,amsfonts}%
\usepackage{amsthm}%
\usepackage{mathrsfs}%
\usepackage[title]{appendix}%
\usepackage{xcolor}%
\usepackage{textcomp}%
\usepackage{manyfoot}%
\usepackage{booktabs}%
\usepackage{algorithm}%
\usepackage{algorithmicx}%
\usepackage{algpseudocode}%
\usepackage{listings}%
\usepackage{orcidlink}

\begin{document}

\title[High-throughput spin-bath characterization of spin-defects in semiconductors]{High-throughput spin-bath characterization of spin-defects in semiconductors}


\author[1,2]{\fnm{Abigail N.} \sur{Poteshman}\orcidlink{0000-0002-4873-4826}}

\author[3]{\fnm{Mykyta} \sur{Onizhuk}\orcidlink{0000-0003-0434-4575}}

\author[2,4]{\fnm{Christopher} \sur{Egerstrom}\orcidlink{0000-0001-8240-7303}}

\author[2]{\fnm{Daniel P.} \sur{Mark} \orcidlink{0000-0002-1231-6016}}

\author[2, 4, 5]{\fnm{David D.} \sur{ Awschalom} \orcidlink{0000-0002-8591-2687}}

\author[2,4]{\fnm{F. Joseph} \sur{Heremans} \orcidlink{0000-0003-3337-7958}}

\author[2,3,4]{\fnm{Giulia} \sur{Galli} \orcidlink{0000-0002-8001-5290}}

\affil[1]{\orgdiv{Committee on Computational and Applied Mathematics}, \orgname{University of Chicago}, Chicago, IL 60637, USA}

\affil[2]{\orgdiv{Materials Science Division}, \orgname{Argonne National Laboratory}, Lemont, IL 60439, USA}

\affil[3]{\orgdiv{Department of Chemistry}, \orgname{University of Chicago}, Chicago, IL 60637, USA}

\affil[4]{\orgdiv{Pritzker School of Molecular Engineering}, \orgname{University of Chicago}, Chicago, IL 60637, USA}

\affil[5]{\orgdiv{Department of Physics}, \orgname{University of Chicago}, Chicago, IL 60637, USA}

\abstract{Detailed knowledge of the local environments of spin-defects in semiconductors, such as nitrogen vacancy (NV) centers in diamond or divacancies in silicon carbide, is crucial for optimizing control and entanglement protocols in quantum sensing and information applications. However, a direct experimental characterization of individual defect environments is not scalable, as spin bath measurements are extremely time consuming. In this work, we address the ill-posed inverse problem of recovering the atomic positions and hyperfine couplings of random nuclei surrounding spin-defects from sparse experimental coherence signals, which can be obtained in hours. To address the challenge to determine the number of isotopic nuclear spins along with their hyperfine couplings, we employ a trans-dimensional Bayesian approach that incorporates \textit{ab initio} data. This approach provides posterior distributions of the numbers, hyperfine couplings, and locations of nuclear spins present in the sample. In addition to enabling high-throughput screening of spin-defects, we demonstrate how this trans-dimensional Bayesian approach can guide experimental design for dynamical decoupling experiments to detect nuclear spins within targeted hyperfine coupling regimes. While the primary focus is on accelerating spin-defect characterization, this Bayesian approach also lays the foundation for digital twin studies of spin-defects, where a virtual model of the spin-defect system evolves in real time with ongoing experimental measurements. Together, the set of tools we designed and applied paves the way for scalable deployment of spin-defects in semiconductors for quantum sensing and information applications. }




\maketitle

\section{Introduction}
\label{intro}

Optically active spin defects in solid-state systems, such as phosphorus donors in silicon \cite{morello2020donor}, divacancies in silicon carbide \cite{castelletto2020silicon}, and nitrogen-vacancy (NV) centers in diamond \cite{rodgers2021materials}, have emerged as promising platforms for quantum information science due to their long coherence times \cite{anderson2022five, kanai2022generalized} and their coupling to local nuclear spin environments \cite{bourassa2020entanglement, bradley2022robust}. While long coherence times are critical for quantum applications, the presence of naturally occurring nuclear spin isotopes in these systems introduces both challenges and opportunities. On the one hand, nuclear spins are one of the main mechanisms of decoherence, necessitating precise characterization of the local environment to optimize control sequences and enhance performance \cite{waeber2019pulse}. On the other hand, these same nuclear spins may serve as auxiliary quantum resources, enabling functionalities such as local quantum registers or quantum memories, provided their hyperfine couplings and spatial configurations are known \cite{dong2020precise, cramer2016repeated, taminiau2012detection, taminiau2014universal}. 

Recent advances have enabled scalable fabrication of optically-active spin-defects in semiconductors (e.g., billions of NV centers in diamond can be created within a few days \cite{marcks2024guiding, horn2024controlled}), prompting a need for high-throughput characterization tools capable of quickly screening large numbers of candidate defects for nuclear spin bath properties for a target quantum information application. While scalable techniques exist to assess optical and spin properties \cite{sutula2023large}, there is currently no efficient approach to determine the local nuclear spin environment of individual defects at scale. High-resolution methods such as correlation spectroscopy provide access to detailed hyperfine and spatial information but require hours of integration time per data point and manual tuning of pulse sequences, rendering them impractical for large-scale deployment \cite{laraoui2013high, abobeih2019atomic}. Machine learning approaches have shown promise for extracting nuclear spin information from coherence data \cite{jung2021deep, varona2024automatic}, but they depend on high-quality, high-resolution datasets that are infeasible to collect in high-throughput settings. 

In this work, we address the ill-posed, trans-dimensional inverse problem of identifying the number, positions, and hyperfine couplings of nuclear spins interacting with a single spin-defect from sparse, noisy coherence data. We focus on a data regime in which experimental coherence signals can be measured rapidly but may be limited in resolution and precision. We refer to this regime as as the sparse, noisy regime, which enables rapid initial screening of candidate defects but poses fundamental challenges for accurate characterization. The inverse problem is ill-posed because the same experimental data may be consistent with multiple distinct nuclear spin configurations, each with different numbers, positions, and hyperfine couplings of nuclear spins.

Existing approaches for extracting information about the nuclear spin environment of single defects from low-resolution data often rely on fitting parameterized models, using off-the-shelf nonlinear optimization methods \cite{vorobyov2022addressing, marcks2024guiding}. These methods require specifying the number of nuclear spins \textit{a priori}, necessitating a separate model selection step, and often ignore physical constraints such as spatial lattice geometry or bandwidth limits. As a result, they may produce unphysical or uninterpretable results. Classical statistical approaches favor lower dimensional models when errors between higher- and lower-dimensional parameterizations are comparable \cite{neath2012bayesian}, but in this setting, there is no physical reason to prefer, for instance, a five-spin nuclear spin configuration over a ten-spin one if both lie within the the expected statistical distribution for a given isotope concentration. 

To achieve efficient and scalable characterization of nuclear spin environments of single defects, we apply a Bayesian framework for recovering nuclear spin environments from sparse coherence data acquired from dynamical decoupling experiments that we introduced in Ref. \cite{poteshman2025transdimensional}. This method performs joint model selection and parameter estimation using reversible jump Markov chain Monte Carlo (RJMCMC) \cite{green2009reversible}, enabling trans-dimensional model selection over the number of nuclear spin, in addition to parameter estimation of the hyperfine couplings. By incorporating lattice parameters and hyperfine couplings calculated from \textit{ab initio} density functional theory (DFT) calculations and enforcing  experimental bandwidth constraints, our approach yields physically meaningful posteriors over nuclear spin configurations. 

Given the ill-posed nature of recovering detailed nuclear spin information from sparse coherence data, our approach does not aim to fully resolve precise hyperfine couplings or nuclear spin locations. Instead, it provides posterior distributions over possible hyperfine coupling and spin location configurations, identifying defects with characteristics that are promising for specific applications. Defects showing favorable hyperfine coupling or nuclear spin distributions can then be selected for further experimental characterization using more time-consuming techniques, while avoiding the allocation of both experimental and computational resources to defects unlikely to meet application-specific requirements.

After providing background in Sec. \ref{sec:background} on the NV center and dynamical decoupling experiments we study here, we begin in Sec. 
\ref{sec:detect_limits} by analyzing the fundamental limits of hyperfine coupling detectability in sparse, noisy regimes, and we use these limits to guide experimental design. In Sec. \ref{sec:application_hybrid_alg}, we then apply a trans-dimensional Bayesian model selection and parameter estimation method that reconstructs the nuclear spin environment by efficiently sampling over nuclear spin configurations with different numbers of nuclear spins. We apply a hybrid MCMC method in two steps: first, in Sec. \ref{subsec:guide_exp_design}, we demonstrate how we can use simulations of nuclear spin baths, within the fundamental limits on detectability, and we show how the hybrid MCMC method can guide experimental design for fast, scalable detection of nuclear spin bath properties; next, in Sec. \ref{subsec:exp_data_results}, we apply the hybrid MCMC method to dynamical decoupling data from ten single NV centers in diamond with a natural abundance of $^{13}$C nuclear spins. We conclude in Sec. \ref{discussion} with a discussion on the capabilities, limitations, and possible extensions of the set of tools we provide. Altogether, our results provide a set of tools to design, analyze, and interpret dynamical decoupling experiments that can be performed rapidly for scalable characterization of nuclear spin environments of spin-defects in semiconductors.

\begin{figure}[t!]
    \centering
    \includegraphics[width=\textwidth]{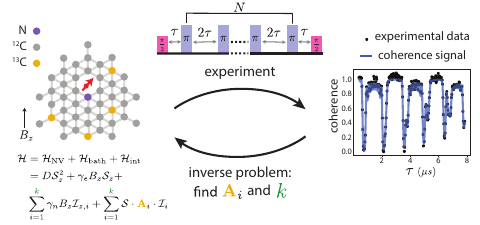} 
    \caption{\textbf{Schematic of inverse problem}. (Left) Representation of a single NV center in diamond with an external magnetic field ($B_Z$) applied along the axis of the NV center, and there are $k$ isotopic $^{13}$C randomly distributed at lattice sites, interacting with the central electronic spin (red) via the hyperfine interaction ($\mathbf{A}$) (see text for the definition of Hamiltonian terms). (Center) Experimental probing by applying $N$ dynamical decoupling $\pi$ pulses with varying inter-pulse spacings ($\tau$) to obtain (Right) a coherence signal. The inverse problem of interest is recovering the number of nuclear spins ($k$) and their hyperfine couplings ($\mathbf{A}$) from a sparse, noisy experimental coherence signal.}
    \label{fig:schematic}
\end{figure}

\section{Background}
\label{sec:background}

\subsection{Hamiltonian description of nuclear spin bath}

We consider a central electronic spin, the nitrogen-vacancy (NV) center in diamond, in an external magnetic field $B_{z}$ interacting with a surrounding bath of nuclear spins, as shown in the schematic in Fig. \ref{fig:schematic}. The dynamics of the system are governed by the Hamiltonian:

\begin{equation}
\mathcal{H} = \mathcal{H}_{\text{NV}} + \mathcal{H}_{\text{bath}} + \mathcal{H}_{\text{int}},
\end{equation}
where $\mathcal{H}_{\text{NV}}$  describes the electronic spin of the NV center, $\mathcal{H}_{\text{bath}}$ accounts for the nuclear spin bath, and $\mathcal{H}_{\text{int}}$ captures the hyperfine interaction between the NV center and the nuclear spins.

The NV center Hamiltonian is given by:
\begin{equation}
\mathcal{H}_{\text{NV}} = D \mathcal{S}_{z}^{2} + \gamma_{e} B_{z} \mathcal{S}_{z},
\end{equation}
where $D $ is the axial component of the zero-field splitting ($D = 2\pi \times 2.87$ GHz for the NV center in diamond), and $\gamma_{e}$ is the gyromagnetic ratio of the NV center. The nuclear spin bath is modeled as a collection of non-interacting spins:
\begin{equation}
\mathcal{H}_{\text{bath}} = \sum_{i=1}^k \gamma_{n} B_{z} \mathcal{I}_{z,i},
\end{equation}
where $\gamma_n$ is the nuclear gyromagnetic ratio. Given the time and energy scales of the sparse, noisy experiments, we neglect interactions among the nuclear spins, since they are relatively weak, on the order of $\gamma_n^2$, compared to electron-nuclear spin interactions. The hyperfine interaction is given by:
\begin{equation}
\mathcal{H}_{\text{int}} = \sum_{i=1}^k \mathcal{S} \cdot \mathbf{A}_{i} \cdot \mathcal{I}_i = \mathcal{S}_{z}(\sum_{i=1}^k A_{\parallel,i} \mathcal{I}_{z,i} + A_{\perp,i} \mathcal{I}_{x,i} )
\end{equation}
where $\mathcal{A}$ represents the hyperfine tensor for each nuclear spin, and since the electronic level splitting is the dominant energy scale, we apply the secular approximation to describe the hyperfine interaction in terms of a parallel ($A_{\parallel}$) and a perpendicular ($A_{\perp}$) component with respect to the externally applied magnetic field $B_z$.

\subsection{Dynamical decoupling experiments}

Dynamical decoupling (DD) sequences, such as the Carr-Purcell-Meiboom-Gill (CPMG) protocol, provide a powerful tool for probing the local spin environment by selectively filtering environmental noise at specific frequencies. These sequences are widely employed in experiments with nitrogen-vacancy (NV) centers in diamond and other solid-state spin qubits to extract information about nuclear spin dynamics and spin-spin interactions \cite{zhao2014dynamical, kolkowitz2012sensing}.

In a CPMG-$N$ experiment, a series of $N$ equally spaced $\pi$-pulses with inter-pulse spacing $2\tau$ are applied to the electronic spin (with $\tau$ time separating the initial and final $\pi$- and $\pi/2$-pulses), effectively modulating its interaction with the surrounding nuclear spin bath. The coherence of the electronic spin, measured as a function of the inter-pulse delay time $\tau$, carries signatures of the nuclear spin bath, including the spectral density of its fluctuations and the spatial configuration of nuclear spins. The modulation function induced by the DD sequence acts as a spectral filter, enhancing the sensitivity to specific nuclear spin precession frequencies that match harmonics of the filter function peaks \cite{alvarez2011measuring}.

A widely used theoretical approach for modeling electronic spin coherence under DD sequences is the cluster-correlation expansion (CCE), which systematically accounts for many-body interactions within the nuclear spin bath \cite{yang2008quantum, onizhuk2021pycce}. The CCE method provides a numerically controlled approximation to the coherence function by truncating the expansion at a given cluster order, where CCE-$n$ includes interactions up to $n$-body nuclear spin correlations. For dilute nuclear spin baths where nuclear-nuclear interactions are weak, the lowest-order term, CCE-1, provides a quantitatively accurate description, capturing the dominant contribution to decoherence from independent nuclear spins \cite{taminiau2012detection}.

In this work, we focus on a regime where nuclear-nuclear interactions can be neglected, allowing for a closed-form analytical expression for the electronic spin coherence. This analytical formula, derived from the semiclassical treatment of the spin bath, is formally equivalent to the CCE-1 approximation and provides an exact description of the coherence decay under CPMG sequences in the absence of nuclear spin interactions for NV centers in diamond \cite{taminiau2012detection}. While CCE-1 is computationally efficient and yields accurate results for coherence calculations for NV centers in diamond, it is not universally applicable to all spin-defect systems. For example, for defects in SiC, coherence functions are well-converged only at the CCE-2 level of theory \cite{seo2016quantum}. The Bayesian framework we employ, however, is general and can incorporate higher-order CCE calculations as required. As such, this framework is applicable to a broader class of spin-defect systems beyond the examples on NV centers we present here.

The coherence $\mathcal{L}$ for a known configuration of $k$ nuclear bath spins $\{s_i\}_{i=1}^k$ for a CPMG-$N$ experiment with inter-pulse spacing $\tau$ performed on an NV center with an external magnetic field $B_z$ is given by:

\begin{equation}
    \mathcal{L}(\{s_i\}_{i=1}^k, \tau, N, B_z) = \frac{1}{2}(1+\prod_{i=1}^kM(A_{\parallel_i}, A_{\perp_i}, \tau, N, B_z))e^{-\frac{\tau}{\lambda}}
    \label{eq:coherence_formula}
\end{equation}

\begin{equation}
    M(A_{\parallel, k}, A_{\perp, i}, t) = (1 - m_{i,x}^2 \frac{(1- \cos{\alpha_i})(1-\cos{\beta})}{1 + \cos{\alpha_i}\cos{\beta} - m_{i,z}\sin{\alpha_i}\sin{\beta}} \sin{\frac{N\phi_i}{2}}^2)
    \label{eq:M_k}
\end{equation} for $\cos{\phi_i} = \cos{\alpha_i}\cos{\beta} - m_{i,z}\sin{\alpha_i}\sin{\beta}$ with $m_{i,z} = \frac{A_{\parallel, i} + \omega_L}{\tilde{\omega}_i}$ and $m_{i,x} = \frac{A_{\perp, i}}{\tilde{\omega}_i}$ for $\tilde{\omega}_i = \sqrt{(A_{\parallel, i} + \omega_L)^2 + A_{\perp, i}^2}$, $\alpha_i = \tilde{\omega}_i t$, and $\beta = \omega_L t$, and $\lambda$ is a decay factor to account for dephasing of the electronic spin due to interactions that are not explicitly modeled here, such as interactions with lattice impurities other than $^{13}$C that may be present in experimental samples or interactions with the distant spin bath, which can cause an overall decay in the coherence signal \cite{taminiau2012detection, jung2021deep}. We find that including the decay factor $\lambda$ to implicitly account for dephasing effects is crucial to ensure that our forward model (Eq. \ref{eq:coherence_formula}) properly describes the coherence experiment in the sparse, noisy experimental data regime. See Supplemental Information Sec. 1 for more details on $\lambda$.

\section{Detection limits imposed by noisy, sparse data}
\label{sec:detect_limits}

\begin{figure}[t!]
    \centering
    \includegraphics[width=0.8\textwidth]{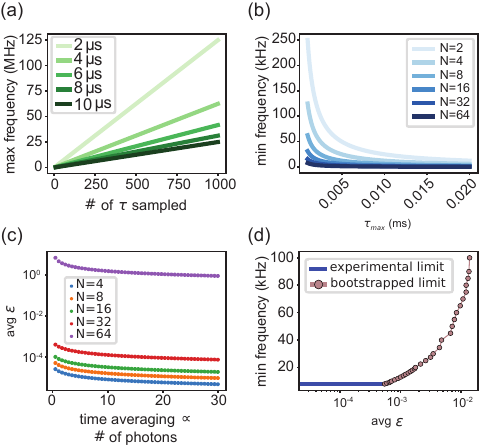} 
    \caption{\textbf{Hyperfine coupling frequency detection limits from frequency sampling limits and simulations} \textbf{(a)} Maximum detectable frequencies as a function of the number of interpulse spacings ($\tau$) sampled, shown for various maximum interpulse spacings ($\tau_{\text{max}}$). \textbf{(b)} Minimum detectable frequencies as a function of $\tau_{\text{max}}$ for different Carr-Purcell (CP) pulse numbers ($N$). \textbf{(c)}  Schematic of average shot noise ($\varepsilon$) per data point for varying numbers of CP-pulses ($N$) with an arbitrary initial signal-to-noise ratio. \textbf{(d)} Minimum detectable hyperfine frequency, calculated as a function of the average $\varepsilon$ per data point, for joint fits of an N=8 and N=16 CP pulse experiment with  $\tau_{\text{max}} = 8 \mu$s, $B_{\text{ext}} = 311$G, and 250 $\tau$ sampled at the natural concentration (1.1\%) of $^{13}$C. The bootstrapped limit is shown for 0.95 confidence of detectability. The minimum frequency limit imposed by the experimental settings (7.8125 kHz).}
    \label{fig:hf_bounds}
\end{figure}

Because we aim to recover hyperfine couplings from experimental coherence signals that are relatively short and sparsely sampled in terms of interpulse spacings $\tau$, the resulting data are inherently noisy due to the photon detection process. This shot noise, combined with the limited frequency resolution dictated by the experimental parameters, constrains the extraction of meaningful hyperfine couplings. To systematically determine the regime where extracted hyperfine couplings remain physically significant rather than artifacts of shot noise, we analyze next the interplay between sensing limits, nuclear spin bath properties, and experimental averaging. These factors collectively determine how the chosen experimental parameters influence the physical significance of the extracted hyperfine couplings.

\subsection{Detection limits imposed by sparsity}
The fundamental sensing limits in dynamical decoupling experiments are governed by the Nyquist-Shannon sampling theorem and the maximum interpulse spacing ($ \tau_{\text{max}} $). The highest resolvable frequency is 

\begin{equation}
f_{\text{max}} = \frac{\text{number of sampled interpulse spacings}}{4 \times \tau_{\text{max}}},
\end{equation} 
while the lowest detectable frequency is limited by the duration of the experiment, given by 
\begin{equation}
f_{\text{min}} = \frac{1}{N \times 2 \times \tau_{\text{max}}}
\end{equation}
where $N$ is the number of dynamical decoupling pulses and the total length of an experiment is $N \times 2 \times \tau$ for a given $\tau$. These limits define the range of frequencies that can be extracted for a fixed magnetic field, $ B_{\text{z}} $, given the window and number of interpulse spacings sampled (Fig. \ref{fig:hf_bounds}a \& b).

For nuclear spin environments, the Zeeman energy splitting induced by $B_{\text{ext}}$ determines the nuclear spins' Larmor precession frequency, thereby affecting the dynamical evolution of the system. Specifically, higher values of $ B_{\text{z}} $ result in faster Larmor precession, effectively compressing the characteristic timescales of the dynamics, while lower values extend these timescales. This relationship establishes a redundancy between $ \tau_{\text{max}} $ and $ B_{\text{z}} $, expressed as $\tau_{\text{max}} \propto \frac{1}{\gamma B_{\text{ext}}}$, where $ \gamma $ denotes the nuclear spins' gyromagnetic ratio.

In this work, we explore the impact of dynamical timescales under fixed $ B_{\text{z}} $ while varying $\tau_{\text{max}}$. This choice reflects the experimental ease of modifying interpulse spacing, which only requires programmatic adjustments to the experimental control sequence, compared to altering $ B_{\text{z}} $, a process that involves physical reconfiguration and fine-tuning of multiple hardware components. While increasing $ B_{\text{z}} $ shortens the timescales, it may also alter effective noise levels and potentially affect resolution limits due to environmental interactions. A detailed investigation of these noise and resolution effects is beyond the scope of this study but represents a direction for future work.

\subsection{Detection limits imposed by shot noise}

Each spin defect acts as a single-photon emitter, limiting the number of spin-sensitive photons that can be detected in a single measurement before the spin state decoheres or reinitializes. As a result, experimental coherence signals are acquired by counting single photons and averaging over repeated measurements, yielding data distributed according to Poisson statistics. In high-throughput settings, the requirement of rapid acquisition imposes a fundamental trade-off between measurement time and data quality. Here we investigate how noise impacts our ability to recover hyperfine couplings from coherence data acquired in the sparse, noisy regime. In particular, we examine the extent to which weakly coupled nuclear spins can be reliably detected. This analysis establishes fundamental limits on the minimum detectable hyperfine coupling strength given a fixed amount of shot noise and informs experimental design choices for scalable characterization.

To systematically determine the crossover threshold at which hyperfine couplings of specific frequencies become undetectable, compared to shot noise for a fixed set of $^{13}$C isotopes surrounding NV centers in experiments, we fixed key experimental parameters. These include the applied external magnetic field ($B_{\text{ext}}$), the number of dynamical decoupling pulses ($N$), the maximum interpulse spacing measured ($\tau_{\text{max}}$), the number of interpulse spacings measured, the amount of shot noise $\varepsilon$, and the isotopic concentration of spins $c$. We then sampled hyperfine coupling constants for atomic positions within 30 \AA\ of the NV center using \textit{ab initio} density functional theory calculations \cite{takacs2024accurate} and using the dipole approximation for atoms between 30 \AA\ and 40 \AA. A 40 \AA\ cutoff leads to well-converged results for the noise regimes and pulse sequences studied here; nuclear spins beyond this distance are so weakly coupled that their contributions are below the detection threshold imposed by shot noise.

Using a specified concentration ($c$), we generated 10,000 random configurations of lattice sites and, for each configuration, calculated the $L_2$ error between the coherence signal $L(t)$ obtained with the full bath and the signal $L'(t)$ resulting from omitting spins with hyperfine couplings below a specified frequency threshold. To assess the uncertainty in the errors for each threshold value, we employed a bootstrapping procedure with 10,000 iterations, resampling the calculated errors to obtain the 0.95 confidence level. Specifically, the 0.95 confidence value corresponds to the error level such that 95\% of spin configurations yield smaller errors than the expected error from the shot noise. The averaged bootstrapped 95\% confidence error was then normalized by the number of time points in the coherence signal (i.e., the number of interpulse spacings). This normalized value represents the experimental noise level for a given hyperfine threshold such that the perturbation induced by nuclear spins with small hyperfine couplings exceeds the noise with 95\% confidence (Fig. \ref{fig:hf_bounds}c).

\subsection{Determining limits on detectability from sparse, noisy experimental data}

By combining fundamental limits on hyperfine frequency detectability with calculations of the crossover regime between intrinsic and external noise, we determined how the minimum detectable hyperfine frequency decreases with experimental averaging. This approach allows us to obtain the minimum detectable hyperfine frequency for a particular experiment with a desired amount of averaging. Notably, for the experimental parameters used in this study ($\tau_{\text{max}} = 8 \mu$s, $B_{\text{ext}} = 311$G, and 250 $\tau$ sampled for CP-8 and CP-16 pulses), we established a lower detectability limit of 7.8125 kHz, below which hyperfine couplings cannot be resolved regardless of experimental averaging (Fig. \ref{fig:hf_bounds}d).

Our combined approach achieves two objectives.  First, given a specific experimental dataset, including the amount of time averaging, we can verify that the extracted hyperfine couplings are meaningful rather than extraneous fitting artifacts. Second, for targeted applications of spin defects in semiconductors, we can design optimal dynamical decoupling experiments. By tailoring the number of time points sampled, the number of pulses measured, the applied magnetic field, the interpulse spacing, and the required time averaging, we can ensure efficient detection of hyperfine couplings within a desired range. This approach paves the way for high-throughput characterization strategies, enabling precise and resource-efficient evaluation of spin-defect properties.

\section{Application of a hybrid MCMC algorithm for experimental design and nuclear spin bath recovery}
\label{sec:application_hybrid_alg}

Having established fundamental constraints on the detectability of hyperfine couplings that hold independently of the specific method used for parameter recovery, we now turn to the application of a concrete inference strategy designed to operate effectively in sparse and noisy experimental regimes. To this end, we employ a hybrid Markov chain Monte Carlo (MCMC) algorithm that integrates several complementary techniques: reverse jump MCMC (RJMCMC), parallel tempering (PT), and continuous and discrete versions of random walk Metropolis-Hastings (RWMH). Each of these components addresses a distinct challenge associated with the high-dimensional and trans-dimensional nature of the inverse problem under consideration. Our approach integrates \textit{ab initio} information about the spin-defect system, including the crystal lattice geometry and hyperfine couplings computed from first principles \cite{takacs2024accurate}. Using this structural and electronic input as a prior, we define a probabilistic model over configurations of nuclear spins on lattice sites and their associated decoherence effects.

To obtain the posterior distribution over spin configurations and parameters, we employ a hybridized Markov chain Monte Carlo (MCMC) approach introduced in Ref. \cite{poteshman2025transdimensional}. The hybridized algorithm is built around a discrete random walk Metropolis-Hastings (RWMH) algorithm, which samples lattice sites via a discrete random walk. Reverse jump MCMC (RJMCMC) addresses the trans-dimensional nature of the inference problem by proposing birth and death moves at lattice sites to vary the number of nuclear spin walkers \cite{green2009reversible}. Within a fixed-dimensional model, parallel tempering (PT) is employed to overcome the non-linear and non-convex structure of the likelihood landscape by running multiple chains of discrete RWMH at different effective temperatures, with exchanges between chains enhancing convergence and avoiding local minima. Finally, continuous RWMH updates are used to obtain posterior distributions for the real-valued relative decoherence scaling factor $\lambda$, which captures spin decoherence contributions beyond the dominant hyperfine couplings. This hybridized MCMC strategy provides a general and robust framework for extracting spin bath parameters across a wide range of experimental conditions and materials systems. Appendix Sec. \ref{sec:method_hybrid_alg} outlines the hyperparameters used in this work.

We demonstrate the utility of this approach in two settings. First, we demonstrate how this hybrid MCMC algorithm can be applied to guide the design of rapid, resource-efficient experiments targeting nuclear spins with hyperfine couplings in a specified range (Sec. \ref{subsec:guide_exp_design}). Second, we demonstrate how this hybrid MCMC approach enables high-throughput screening of spin-defect samples for desirable nuclear spin bath properties from dynamical decoupling data (Sec. \ref{subsec:exp_data_results}).

\subsection{Guiding experimental design}
\label{subsec:guide_exp_design}

\begin{figure}[t!]
    \centering
    \includegraphics[width=0.8\textwidth]{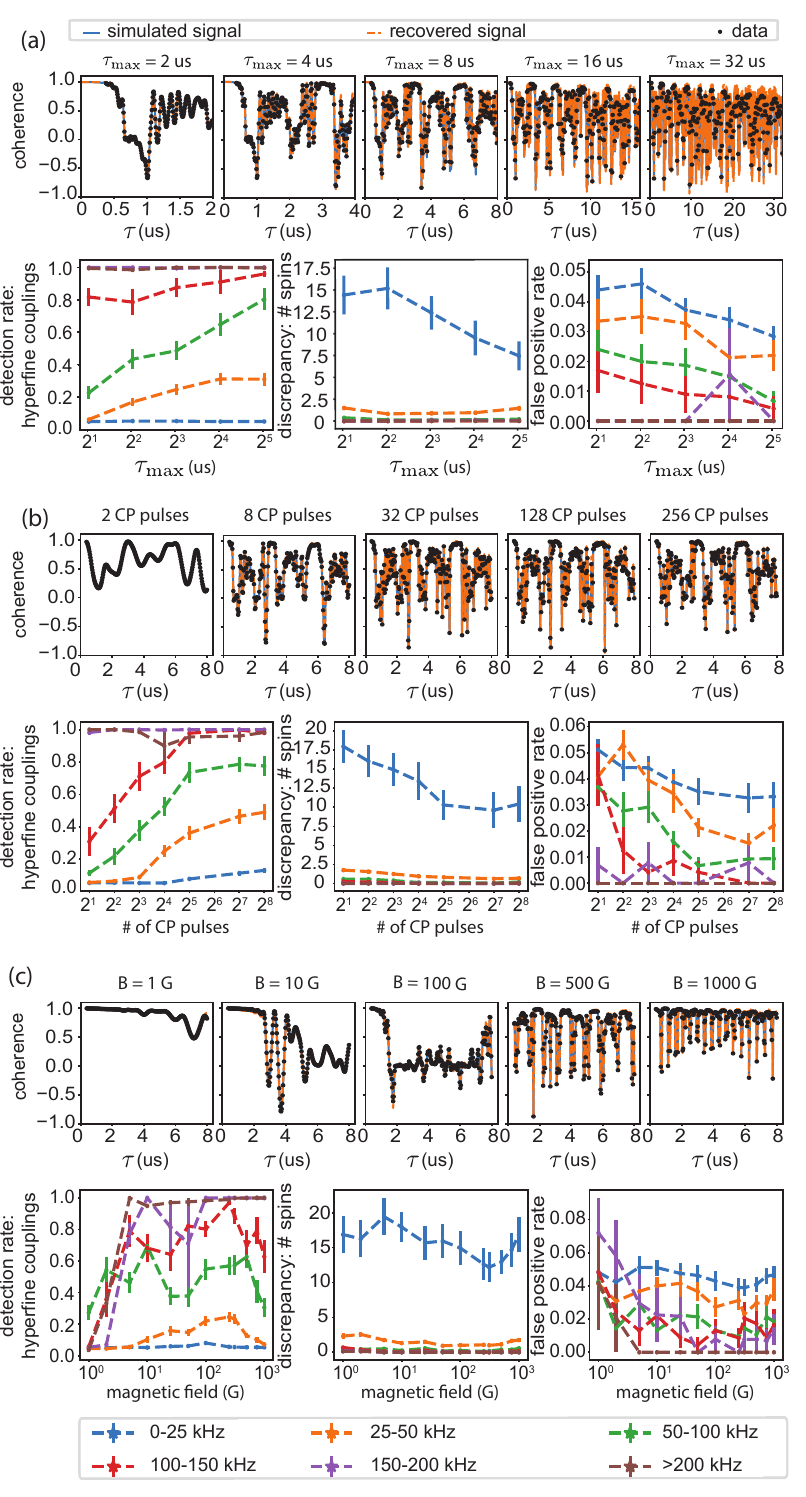} %
    \caption{\textbf{Accuracy of hybrid MCMC algorithm for recovering hyperfine couplings from simulated data.} 
    We applied our hybrid recovery method (see text) to simulations with 0.001 shot noise per data point, 250 $\tau$ sampled, a 0.008 ms longest $\tau$, 16 CP pulses, and 311 G magnetic field, and varied each one of the parameters independently for (a) maximum interpulse length $\tau_{\text{max}}$, (b) the number of CP pulses, and (c) magnetic field. For each of the parameters varied, we plot examples of simulated and recovered signals based on the data and detection rates, discrepancies in the number of recovered versus simulated spins, and false positive rates for the hybrid MCMC method averaged over random nuclear spin configurations consisting of 5 to 20 nuclear spins (for a total of 16 simulated nuclear spin configurations). 
    Results are plotted by groupings of nuclear spins based on hyperfine magnitudes ($\sqrt{A_{\parallel}^2 + A_{\perp}^2}$) (see bottom colorbar).} 
    \label{fig:simulation_results}
\end{figure}

To determine which experiments are most effective for detecting nuclear spins with hyperfine couplings in a targeted range, we use simulations to systematically evaluate the performance of our hybrid MCMC recovery method across varied experimental conditions. This approach enables experimentalists to design and select protocols that maximize recovery success for their detection goals, even in the presence of constraints imposed by experimental setups, such as fixed magnetic fields, limited pulse numbers, or bounded total acquisition time.

For each simulation, we generate synthetic coherence signals based on model nuclear spin baths and simulate measurements under different experimental configurations. These configurations include variations in the number of Carr-Purcell (CP) pulses, the maximum interpulse spacing ($\tau$), and the resolution of $\tau$-sampling. By averaging recovery statistics across multiple simulated spin baths, each with a different number of simulated nuclear spins (between 5 and 20), we assess how well each experimental setting enables accurate inference of spin number and hyperfine parameters. Key performance metrics include the recovery rate of individual spins (i.e., how frequently each true spin appears in the posterior), the discrepancy between the inferred spin number (the modal number of spins in the posterior distribution) and the number of simulated spins, and the false positive rate of the nuclear spins in the modal spin configuration of the posterior distribution. 

Application of the hybrid MCMC method to simulated data demonstrates that the recovery fidelity depends on the interplay between experimental parameters. In particular, we find that hyperfine parameter recovery is correlated with the strength of the hyperfine coupling magnitude across all experimental regimes, where nuclear spins with stronger hyperfine couplings are recovered more accurately than more weakly coupled nuclear spins (see Fig. \ref{fig:simulation_results}). In particular, we find that there is a saturation point beyond which increasing the length of the experiment (maximum $\tau$) and increasing the number of dynamical decoupling pulses yields little additional benefit in terms of recovery, suggesting a regime of diminishing returns with increased experimental resources. For example, we observe a plateau in the detection rate of hyperfine couplings in for hyperfine couplings between $25-100$ kHz for greater than $2^5$ CP pulses, indicating that the recovery of hyperfine couplings is limited not by the number of pulses but by the length of longest interpulse space sampled or the number of time points sampled (see Fig. \ref{fig:simulation_results}). We emphasize that this regime of diminishing returns is a result of the interdependence of multiple experimental parameters (see Sec. 2 of the Supplementary Materials for more details on the impact of covarying experimental parameters on recovery rates) and can thus help experimentalists avoid inefficient dynamical decoupling sampling strategies.

Overall, these simulations enable targeted experimental planning: given a specific detection goal (e.g., recovering hyperfine couplings in the 10–50 kHz range) and hardware constraints, one can identify experimental protocols that will offer the highest likelihood of successful recovery using this hybrid MCMC method. Furthermore, experimental constraints such as fixed magnetic fields or pulse limits can be explicitly incorporated into the simulations, allowing users to determine what is achievable within their setup and to identify tradeoffs (e.g., more averaging vs. more resolution) before performing experiments. This capability is especially valuable in resource-constrained or high-throughput settings. See Appendix Sec. \ref{simulation_results} for further detailed analysis of these simulations.

\subsection{Application to experimental data}

\begin{figure}[h!]
    \centering
    \includegraphics[width=0.8\textwidth]{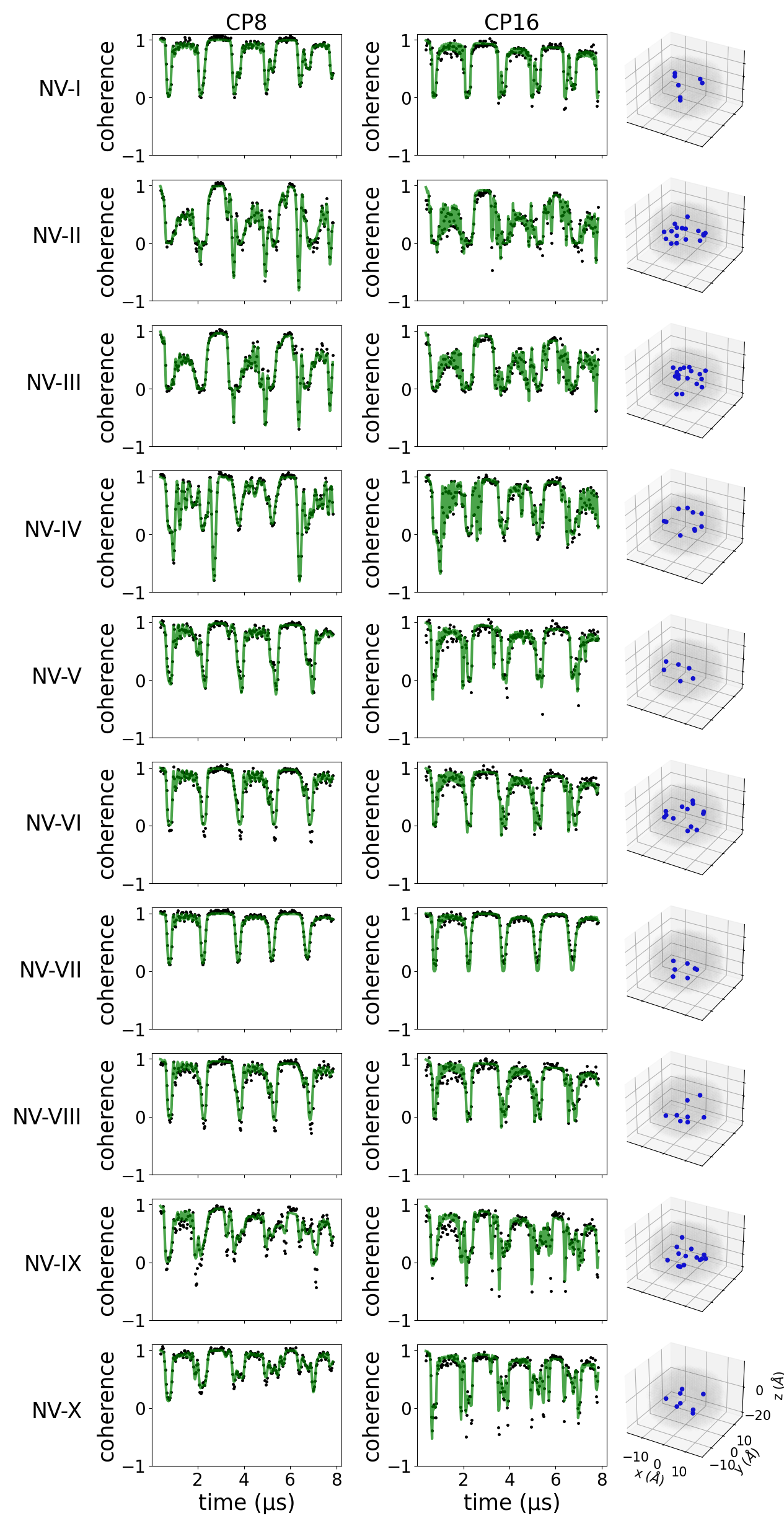} %
    \caption{Experimental coherence signal data for 10 different nitrogen-vacancy (NV) centers under CP8 (left column) and CP16 (middle column) pulse sequences, overlaid with simulated coherence signals (green) generated from the modal spin configuration of the posterior distribution of nuclear spin configurations generated by our hybrid method to extract nuclear spin environments from experimental data. Each row corresponds to a distinct NV center. The right column shows the spatial distribution of nuclear spins in the modal configuration for each NV, plotted relative to the NV center (at the (0, 0, 0) position). Only lattice positions of nuclear spins with hyperfine magnitudes $> 25$ kHz are plotted. The hybrid method jointly analyzes CP8 and CP16 datasets to infer spin environments that best match the experimental data.}
    \label{fig:exp_data}
\end{figure}

 \begin{figure}[h!]
    \centering
    \includegraphics[width=\textwidth]{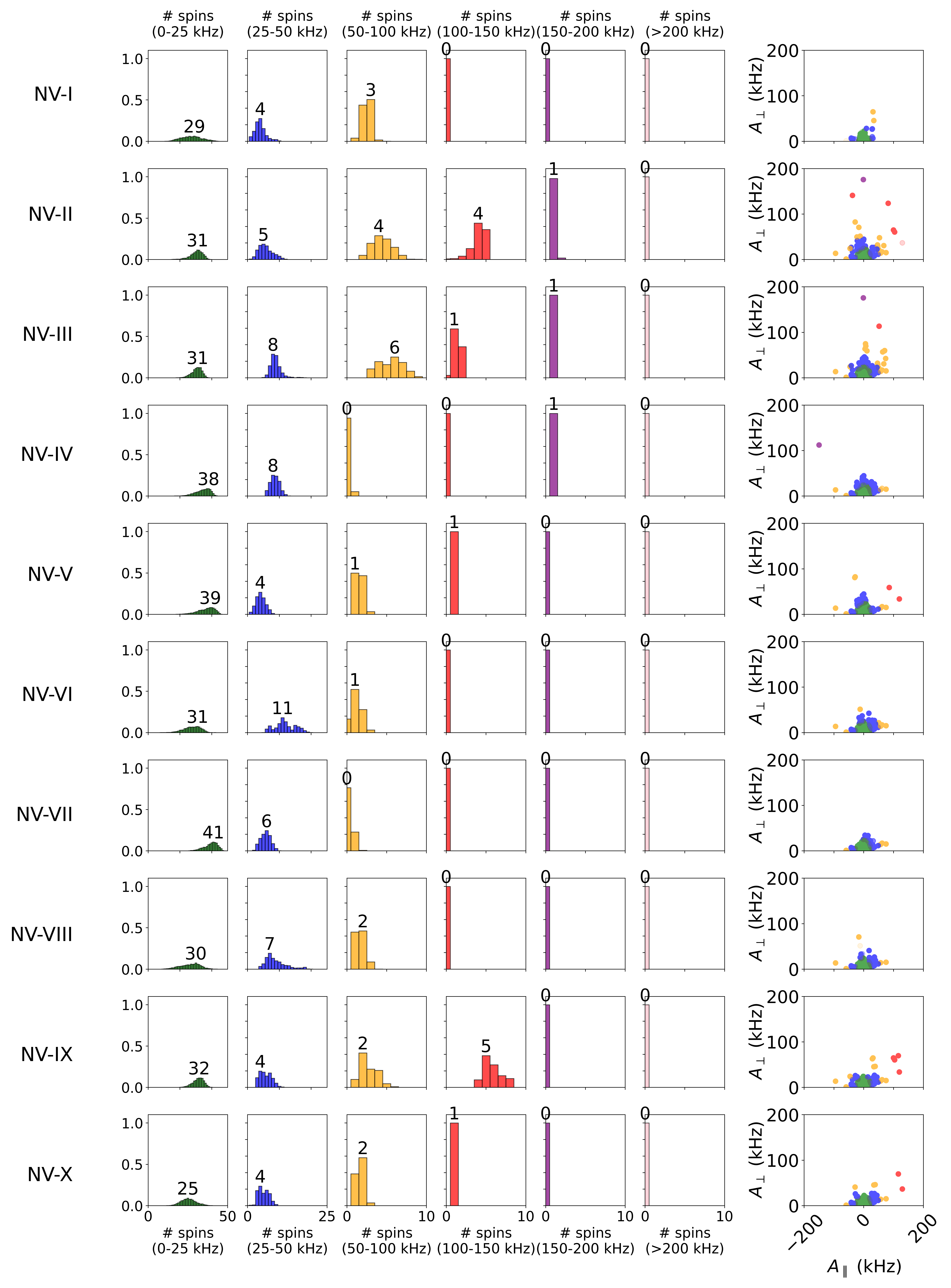} %
    \caption{Posterior distributions of the number and hyperfine couplings of spins from the hybrid algorithm applied to ten NV centers in diamond, plotted by magnitude ($\sqrt{A_{\perp}^2 + A_{\parallel}^2}$) of hyperfine coupling.}
    \label{fig:exp_data_k}
\end{figure}

\label{subsec:exp_data_results}
We further demonstrate the utility of this hybrid MCMC approach for high-throughput charaterization of spin-defects in semiconductors by applying it to coherence data obtained from ten distinct NV centers in diamond. The experimental parameters were selected to include two coherence signals—CP8 and CP16—each comprising 250 time points, with interpulse spacings up to 0.008 ms under a magnetic field of 311 G. Experimental noise was quantified using standard error propagation formulas, incorporating contributions from Poisson noise and photon counts during the transformation of raw photon count data into coherence data (see Supplementary Information, Sec. 3, for details).

For each NV center, we obtained posterior distributions of spin configurations that have a high likelihood of producing the observed coherence signal data. Coherence signals generated from the modal spin configuration in these posterior distributions as well as the corresponding lattice positions of the modal spin configuration are shown in Fig. \ref{fig:exp_data}. Posterior distributions of the hyperfine couplings and numbers of spins in each sample are shown in Fig. \ref{fig:exp_data_k}.

Based on the sensitivity analysis detailed in the previous section, we find that hyperfine couplings greater than ~25 kHz can be reliably recovered. Lower-frequency couplings were less well-constrained, consistent with the limits of the experimental noise and hyperfine thresholds. For certain NV centers, the algorithm identified spin configurations that effectively recreate the experimental data. Notably, NV centers II, III, IV, V, IX, and X exhibit key features—such as a small number of strongly coupled spins—that make them particularly relevant as auxiliary qubits for quantum information applications, including local quantum memory units (Fig. \ref{fig:exp_data_k}).

In contrast, NV centers I, XI, XII, and XIII lack strongly coupled spins and are therefore less promising for quantum information applications (Fig. \ref{fig:exp_data_k}). These NVs could potentially be deprioritized in future experimental studies, allowing resources to be reallocated toward further characterization of more promising candidates.

Certain samples, such as NV-IX, presented systematic underfitting in their coherence signal reconstruction. This suggests the presence of strongly coupled spins not captured by the DFT-calculated hyperfine couplings. Possible explanations include the presence of substitutional nitrogen defects (P1 centers), interactions with nearby NV centers, or other lattice distortions that deviate from the idealized DFT model.

\section{Discussion}
\label{discussion}

Bayesian frameworks offers distinct advantages for characterizing quantum systems compared to machine learning (ML)-based approaches in the sparse data regime. While ML methods excel at pattern recognition and high-dimensional data processing, they often lack interpretability and explicit uncertainty quantification \cite{scorzato2024reliability}. In contrast, Bayesian methods naturally provide posterior distributions that quantify uncertainty, enabling a probabilistic understanding of the recovered spin environment. This interpretability is crucial for experimental design and decision-making, where understanding the confidence in recovered parameters can directly impact subsequent measurements and analyses.

The Bayesian framework introduced in this work offers a practical strategy for optimizing experimental design. By providing probabilistic estimates of spin configurations and their uncertainties, our approach enables experimentalists to identify measurement regimes that maximize information gain while minimizing acquisition time. In addition, the framework proposed in our work lays the groundwork for implementing digital twin approaches to quantum spin systems, where simulations continuously inform and are informed by experimental data. Digital twins have been successfully applied in areas such as photovoltaic characterization \cite{luer2024digital} and quantum dot synthesis \cite{epps2020artificial}, and similar strategies could enhance defect-based quantum technologies. Specifically, the method presented here could enable real-time data assimilation between a physical spin defect and a virtual model of its spin environment. The digital twin could incorporate additional spin dynamics using either high-fidelity ab initio simulations \cite{park2022predicting} or efficient surrogate models \cite{nyshadham2019machine}, supporting a closed-loop system where experiments refine the model, and the model, in turn, guides subsequent measurements.

Despite its strengths, the current implementation has limitations tied to the underlying physical models. The forward model relies on the cluster correlation expansion (CCE) to compute coherence signals from proposed nuclear spin baths, which accurately captures short-range interactions but becomes less reliable for long-range couplings \cite{zhang2020cluster}. While higher-order CCE can, in principle, account for more complex many-body correlations and long-range effects, the computational cost grows combinatorially with cluster size, making it impractical for large-scale inference problems. 

Additionally, the likelihood model assumes fixed hyperfine couplings derived from a single density functional theory (DFT) calculation on an idealized lattice, neglecting uncertainties due to lattice imperfections, unknown strain, or deviations in defect isolation that may arise when processing samples at scale. This idealized treatment excludes the influence of nearby paramagnetic impurities, such as P1 centers, or other types of lattice defects, which can significantly alter the coherence dynamics \cite{park2022decoherence}. However, if the presence of, e.g. P1 centers is known from experiments, these could eventually be included in large scale DFT calculations.

In summary, we present a set of computational tools for scalable spin-defect characterization, with applications ranging from experimental data interpretation to the design of targeted experimental protocols. While we provide simulations and applications of NV centers in diamond, the approach we present is broadly applicable to other defect systems, such as phosphorus donors in silicon and divacancies in silicon carbide, as well as to broader classes of Hamiltonian learning problems. By addressing the challenges posed by high-throughput defect fabrication and the constraints of noisy experimental regimes, our work provides a pathway toward efficient integration of spin defects into quantum technologies.

Several avenues for future work can address these limitations and extend the applicability of this approach. First, this method could be extended to other systems, such as divacancies in SiC \cite{anderson2022five} or phosphorus donors in silicon \cite{mccallum2021donor}. A more robust likelihood model incorporating uncertainty in DFT-derived hyperfine couplings and lattice distortions could further improve accuracy. We could also reframe the problem to find the shortest dynamical decoupling experiment given a desired uncertainty bound for a target regime of hyperfine frequencies. Finally, extending the model to include full Hamiltonian learning, accounting for nuclear spin interactions or ensembles of defects such as NV-NV \cite{maile2024performance} or NV-P1 systems \cite{park2022decoherence}, represents an exciting challenge for future studies. These advancements would enhance the framework’s predictive power and utility across a wide range of quantum systems.

\backmatter

\bmhead{Data \& code availability}
Data and supporting code that support this study will be made available through github: \texttt{https://github.com/pabigail/rjmcmc-hyperfine-recovery}.

\bmhead{Supplementary information}
There is supplementary information.

\bmhead{Acknowledgements}
We thank Daniel Sanz-Alonso, Benjamin Pingault, and Jiefei Zhang for useful discussions. A.N.P. acknowledges support from the DOE CSGF under Award Number DE-SC0022158. The computational and experimental work was primarily supported by the Midwest Integrated Center for Computational Materials (MICCoM) as a part of the Computational Materials Sciences Program funded by the U.S. Department of Energy (M.O., C.E., F.J.H., and G.G.) with additional experimental support from Q-NEXT, a U.S. Department of Energy Office of Science National Quantum Information Science Research Center under Award Number DE-FOA-0002253 (D.M. and D.D.A.). This research used resources of the University of Chicago Research Computing Center.

\appendix
\renewcommand{\thefigure}{A\arabic{figure}}
\setcounter{figure}{0}  

\section{Hybrid MCMC algorithm for recovery of local nuclear spin environment}
\label{sec:method_hybrid_alg}
Using hyperfine couplings for all lattice sites near the central electronic spin, which were calculated using density functional theory (DFT) in Ref. \cite{takacs2024accurate}, we use the hyperfine frequency detection limits to determine the number and location of lattice sites that correspond to the hyperfine coupling frequencies within the detection limits. We use a cut-off point of $s_{\text{max}} = 50$ spins as the maximum number of spins considered for the NV center in diamond based on a dimensional analysis of the expected number of spins for the lattice given a natural abundance of $^{13}$ C within the cut-off radius of $40$ \AA. We randomly initialize a random number of walkers between 1 and $s_{\text{max}}$ for 5 ensembles of walkers. 

We then iterate through a hybrid MCMC algorithm that cycles through 50 steps of reverse jump Markov chain Monte Carlo (RJMCMC), 100 steps of parallel tempering (PT), and 25 steps of random walk Metropolis Hastings (RWMH) for 25,000 total steps. For pseudocode for each algorithm, see \cite{poteshman2025transdimensional}. We consider the first 5,000 steps as burn-in, so the posterior distributions consist of the spin configurations and $T_2$ values corresponding to the last 20,000 steps, and spin configurations and $T_2$ values are combined across ensembles. Unless otherwise specified, we used $R_{\text{spin}} = 5 \AA$, $R_{T_2} = 0.05$, $\sigma^2 = 0.1$, 10 strands for the PT algorithm, and $\beta = 2^{-n}$ for strand $n \in [0, \dots \text{num strands}-1]$ for the PT algorithm. For more details on hyperparameters, see Sec. 3 of the Supplementary Information. While these hyperparameters are not necessarily optimal, they reliably result in convergence to stationary posterior distributions.

\section{Guiding experimental protocols for targeted recovery of nuclear spins from dynamical decoupling data}
\label{simulation_results}
We provide a detailed analysis of the simulation results from Sec. \ref{subsec:guide_exp_design} and from Fig. \ref{fig:simulation_results}.

\subsection{Maximum interpulse spacing}
Increasing the maximum interpulse spacing ($\tau_{\text{max}}$) allows observation of a greater number of Larmor periods, and we observed that extended $\tau_{\text{max}}$ increases recovery rates of hyperfine couplings for spins greater than $25$ kHz (Fig. \ref{fig:simulation_results}). To just resolve the number of spins in the system, however, we observe that discrepancy for the number of recovered spins compared to the number of simulated spins is stable and accurate (close to zero) for spins with magnitudes $> 25$ kHz, and we find that extending $\tau_{\text{max}}$ results in a decreased discrepancy between the number of simulated and the number of recovered spins. This suggests that it is crucial to extend the length of the longest interpulse spacing to accurately resolve hyperfine couplings, especially for weakly coupled spins ($< 25$ kHz), but shorter experiments (on the order of $\tau_{\max} \approx 0.002$ ms for a $N=16$ pulse CP experiment) may suffice for recovering just the number of spins with magnitudes $> 25$ kHz. 

\subsection{Number of CP pulses}
When varying the number of CP pulses, we observed that recovery rates stabilized after approximately $2^5$ pulses for recovery rates of both hyperfine couplings and number of spins (Fig. \ref{fig:simulation_results}). This plateau indicates that, after this point, increasing the number of pulses does not improve recovery rates unless the sampling resolution is increased to match the higher pulse count (see Supplementary Information Sec. 3 for more information on the interdependence of experimental paramters on hyperfine coupling rates). This finding highlights the importance of optimizing the sampling resolution to fully exploit the information content provided by a higher pulse count.

\subsection{Magnetic field}
Finally, we explored the effect of varying the magnetic field strength on recovery performance. Lower magnetic fields (below 100 G) led to a significant drop in detection capability, with both hyperfine coupling recovery and the number of spins detected suffering considerably (Fig. \ref{fig:simulation_results}). However, as the magnetic field strength increased, both the recovery rates for hyperfine couplings and the number of spins showed modest improvements, and the false positive rate decreased slightly. This suggests that higher magnetic fields improve the quality of the coherence signal and enhance the recovery performance, though the improvements are not dramatic beyond $\sim 500$ G for these sets of experimental settings.

\vspace{0.25cm}

\bibliography{ref_main}

\end{document}


\maketitle
\tableofcontents

\section{Relative decoherence parameter}

\begin{figure}[ht]
    \centering
    \includegraphics[width=\textwidth]{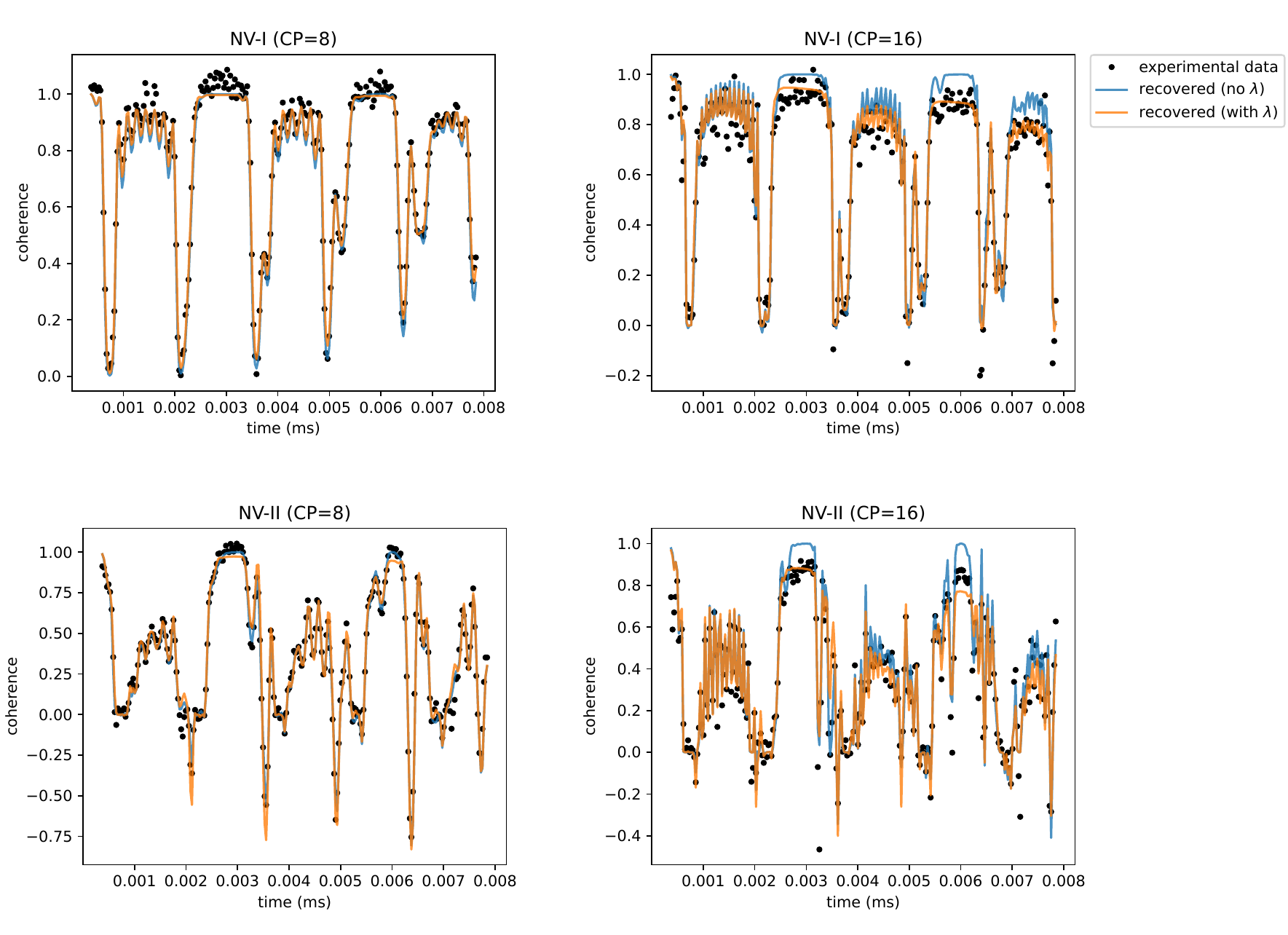} 
    \caption{\textbf{Comparison of coherence signal reconstruction with and without the relative decoherence parameter $\lambda$}. The top row shows experimental data for NV-I with 8 and 16 pulses (CP=8, CP=16); the bottom row shows corresponding data for NV-II. In each panel, black dots denote experimental coherence data, while the blue and orange curves show reconstructed signals derived from the modal spin configuration of the posterior distribution. The blue curves correspond to reconstructions where the relative decoherence parameter $\lambda$ is held fixed at 1, while the orange curves result from inference where $\lambda$ is treated as a free parameter and jointly inferred along with the nuclear spin bath.}
    \label{fig:with_without_lambda}
\end{figure}

In spin-based sensing experiments, decoherence arising from interactions with the surrounding environment becomes increasingly significant at longer evolution times. These effects are particularly relevant in sequences with a large number of dynamical decoupling pulses, such as Carr-Purcell (CP) sequences with 16 or more pulses, where the total experiment duration is extended and the NV center has more opportunity to interact with unmodeled or weakly coupled environmental spins, lattice phonons, or other sources of noise.

As shown in Fig. \ref{fig:with_without_lambda}, neglecting decoherence from sources other than the explicitly-modeled magnetic noise arising from the nuclear spin bath leads to systematic overestimation of the coherence signal at larger values of the pulse interval $\tau$, especially for CP=16. Reconstructions that do not account for these effects—such as those performed with a fixed $\lambda$ parameter—fail to capture the experimentally observed decay. In contrast, incorporating decoherence phenomenologically by inferring an effective regularization parameter $\lambda$ results in reconstructed signals that more accurately reflect the measured coherence, suggesting that this approach can implicitly account for unresolved environmental contributions.

This highlights the importance of including or compensating for decoherence effects when interpreting long-time experimental data, particularly in cases where the model does not explicitly include all sources of environmental coupling.

\section{Interdependence of experimental parameters on hyperfine coupling recovery rates}
\begin{figure}[ht]
    \centering
    \includegraphics[width=\textwidth]{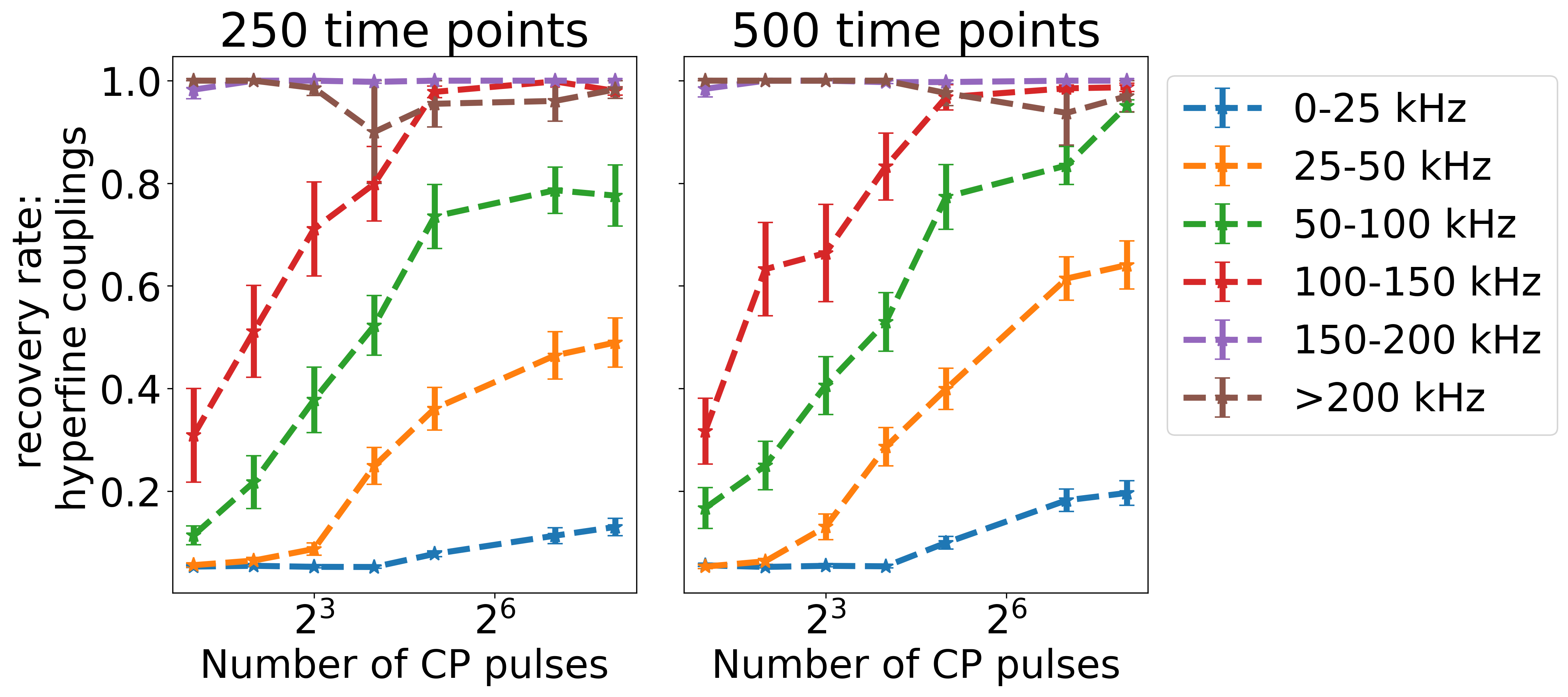} 
    \caption{\textbf{Interdependence of recovery rates on experimental settings.} Recovery rate of hyperfine couplings as a function of the number of CP pulses for two different sampling resolutions: 250 time points (left) and 500 time points (right). Data are grouped by hyperfine coupling strength, ranging from 0–25 kHz to $>$200 kHz. Settings not specified are the same as Fig. 3 of the main text. } 
    \label{fig:interdependence}
\end{figure}

Each CP pulse sequence defines a specific filter function that interacts with the spectral noise, shaping how effectively different frequency components of the noise are suppressed \cite{biercuk2011dynamical}. As the number of pulses increases, the filter function typically broadens and shifts, enabling better suppression of low-frequency noise and leading to an extension of the system's coherence time. For a fixed maximum interpulse spacing, simulations show that recovery rates are comparable across hyperfine couplings of all magnitudes when comparing two sampling rates—250 and 500 time points—up to approximately $2^5$ CP pulses (Fig. \ref{fig:interdependence}). Beyond this point, however, the recovery rate for the 250-time-point sampling plateaus, indicating that the additional information encoded by higher numbers of CP pulses cannot be accessed due to insufficient resolution. In contrast, at 500 time points, the recovery rate continues to improve even beyond $2^5$ CP pulses, suggesting that the higher resolution allows for better extraction of information provided by the increased number of pulses. Additionally, increasing the number of CP pulses tends to extend the system's decoherence time by more effectively suppressing noise, which can further enhancing the recovery of hyperfine couplings (see main text, Sec. ). Thus, it is important to optimize sampling resolution, the number of CP pulses, and the maximum interpulse spacing simultaneously, rather than treating each factor independently, to achieve the best performance in recovering hyperfine couplings.

\section{Experimental error and uncertainty}
\begin{figure}[ht]
    \centering
    \includegraphics[width=0.5\textwidth]{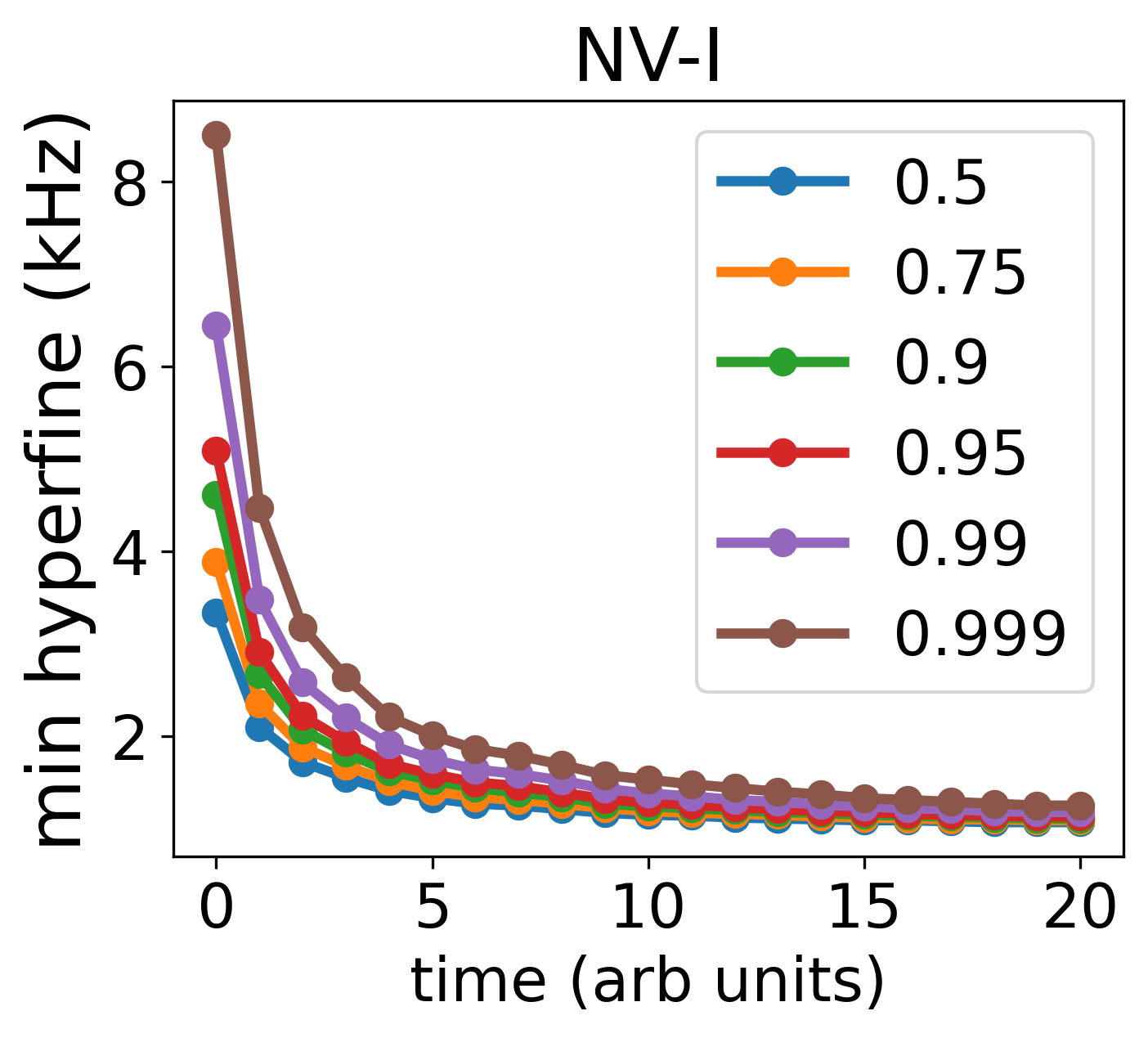} 
    \caption{\textbf{Minimum detectable hyperfine coupling as a function of time averaging}. The time averaging corresponds to a number of raw photon counts that can be used to determine the average uncertainty per data point $\varepsilon$, which can be used as the threshold for the crossover regime in which small hyperfine couplings cannot be reliably distinguished from noise, given a specified statistical confidence. }
    \label{fig:hf_func_time}
\end{figure}

To reduce the effects of fluctuations among the data and calculate uncertainties, the raw photon counts from the bright ($p_z$) and dark ($m_z$) channels are normalized: $ I_\text{norm} = \frac{p_z - m_z}{p_z + m_z}$. The coherence is computed from the normalized bright and dark channels, and we propagate uncertainties based on standard error propagation formula for each manipulation of the raw data into coherence signals. We take the uncertainty of each channel to be the square root of the number of photons detected. As an increasing number of photons are detected, the uncertainty per data point decreases (Fig. 1c in the main text), and as a result the minimum detectable hyperfine frequency decreases as an increased number of photons are collected, as shown in Fig. \ref{fig:hf_func_time} for different values of statistical confidence.

Although the minimum hyperfine detection limit for the experimental data in the main text is limited by the total duration of the experiment rather than by shot noise, we report the average shot noise $\varepsilon$ per data point for the ten samples of NV centers we applied this hybrid algorithm recovery approach to in the main text calculated from the photon counts from the raw experimental data in Table \ref{tab:exp_data_uncertainty}.

\begin{table}[ht]
    \centering
    \begin{tabular}{|c|c|c|c|c|}
        \hline
        \textbf{NV} & 
        \makecell{Avg. Exp. Uncertainty \\ (CP=8)} & 
        \makecell{Avg. Exp. Uncertainty \\ (CP=16)} & 
        \makecell{Lower Hyperfine Threshold \\ from bootstrapping (kHz)} \\ 
        \hline
        I & 0.00266 & 0.00627 & 1.15  \\
        II & 0.00211 & 0.00470 & 1.12  \\ 
        III & 0.00279 & 0.00572 & 1.14  \\ 
        IV & 0.00301 & 0.00620 & 1.15  \\ 
        V & 0.00439 & 0.00850 & 1.22  \\ 
        VI & 0.00369 & 0.00819 & 1.20  \\ 
        VII & 0.00154 & 0.00128 & 1.04 \\ 
        VIII & 0.00495 & 0.00997 & 1.25  \\ 
        IX & 0.00290 & 0.00657 & 1.16  \\ 
        X & 0.00257 & 0.00531 & 1.13 \\ 
        \hline
    \end{tabular}
    \caption{\textbf{Average uncertainty per data point}. Columns corresponding to experimental uncertainties are calculated from photon counts from raw experimental data, and the lower hyperfine threshold is obtained numerically based on the averages of the experimental uncertainties.}
    \label{tab:exp_data_uncertainty}
\end{table}

\section{Comparing posterior distributions to \textit{ad hoc} fitting procedure}
\begin{figure}[ht]
    \centering
    \includegraphics[width=\textwidth]{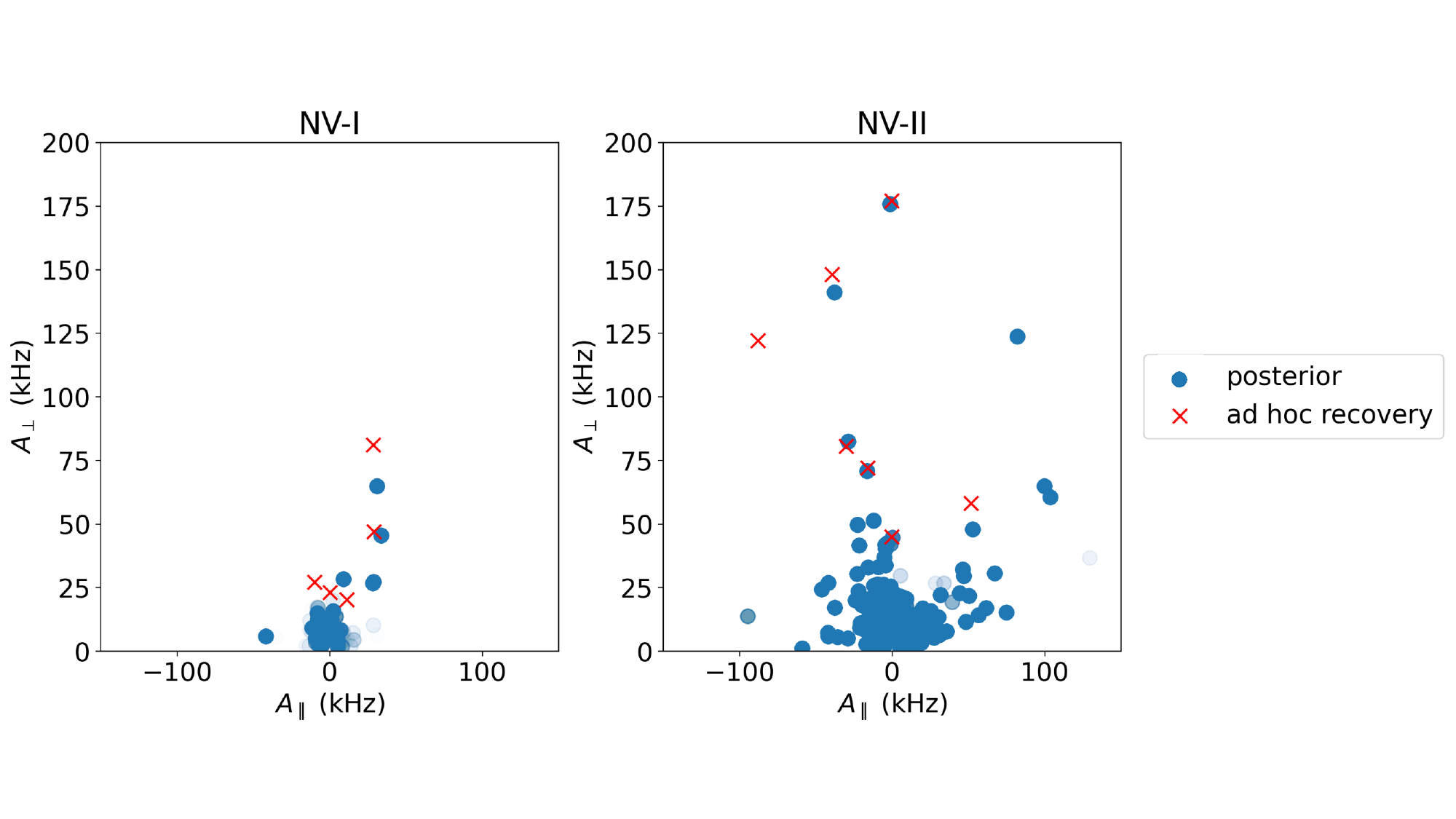} 
    \caption{\textbf{Comparing posterior distributions to ad hoc fitting procedure} Hyperfine coupling frequencies obtained using an \textit{ad hoc} fitting method in \cite{jerger2023detecting} are overlaid on the posterior distributions obtained in this work.}
    \label{fig:compare_adhoc}
\end{figure}

NV-I and NV-II in the main text are the same samples as NV-A and NV-B in \cite{jerger2023detecting}, in which hyperfine couplings and nuclear spins were obtained in an \textit{ad hoc} manner by iteratively fitting different numbers of spins with different hyperfine couplings to the data. We find agreement between the hyperfine coupling parameters of one of the five recovered spins for NV-I and four of the seven recovered spins for NV-II (see Fig. \ref{fig:compare_adhoc}).

\bibliography{ref_supp}